# Enabling Software Defined Optical Networks

Deven Panchal
Student, Dept. Of Electrical and Computer Engineering,
Georgia Institute Of Technology, Atlanta, USA.
devenrpanchal@gatech.edu

**Abstract:** *This paper gives an overview of Software Defined Optical Networks (SDON's) and how they can be implemented. It traces the evolution of Optical networks upto GMPLS and traces the idea of SDN and builds upto OpenFlow. The paper explores the need for SDON's and explains what a SDON solution could look like, including the hardware. It also seeks to explain how OpenFlow could be used as a part of this solution to overcome the limitations of GMPLS.*

## 1 Introduction

Network traffic today is growing due to distributed data centers, machine to machine communications, big data, real time streaming etc. all of which not only involve the transfer of large amount of data, but also have different traffic patterns. Service Providers and data center operators today want to increase capacity on their carrier networks without increasing the cost per bit or the cost that goes into buying network hardware. They want a solution which can provide flexibility i.e. direct customer control over the network resources and also disaggregation i.e. reduced dependence on a single networking hardware vendor. In other words, they want a network with high capacity and which yields low CAPEX and OPEX. Also, because of the rise of dynamic cloud services, service provider want to transform to a DevOps model of rolling out new services.

Most of these fat networks are optical networks because of the fact that fiber medium has the highest capacity. 100 Gbps per channel WDM systems are being deployed globally and superchannels with 400 Gbps or 1 Tbps have been demonstrated [4]. Also, single fiber capacity has exceeded 100 Tbps [2, 3] and this can be increased by spatial division multiplexing [1]. Optical circuits are scalable and also consume lesser power due to passive signal processing. Also, optical components like reconfigurable add drop multiplexers (ROADM's) are much cheaper compared to layer routers. Due to this, optical switching at layer 0 is cheaper than layer-3 IP routing. Fig 1 illustrates this idea in what is called as Verizon's Inverted Triangle.

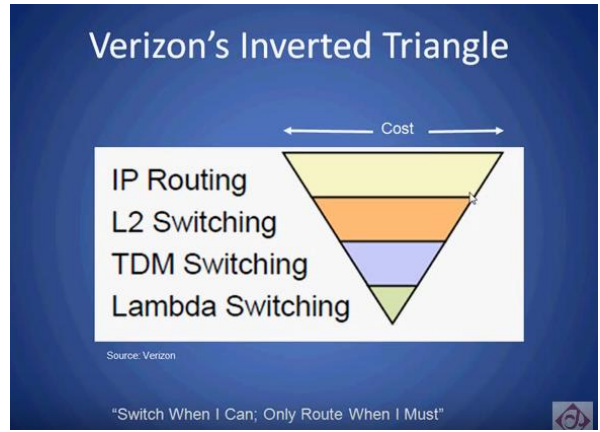

Fig 1: Verizon's Inverted Triangle

The one solution which promises to solve these problems for the network operators is Software defined Networking (SDN). However the SDN technology which was developed for IP based networks cannot be directly transferred to and implemented in circuit switched optical networks. Hence some innovations and extensions are required for it to suit optical networks and the resulting networks are referred to as Software Defined Optical Networks (SDONs).

In addition to solving the aforementioned problems, software defined optical networking has other benefits too. It can lead to increased revenue generation because it can maximize throughput, provides ability to offer networking-as-a-service, can provide on-demand data center connections and can help the network operator implement profitable business models easily. It can also help reduce costs because it aims to achieve global optimization of resources and higher service availability along with better network-wide energy efficiency and all this with fewer staff requirement.

**2 SDN and OpenFlow**

The traditional architecture of the network as shown in Fig 2 (a), consists of interconnections between network devices having specialized hardware, specialized control plane and specialized features that vary from vendor to vendor, making the devices vertically integrated, closed and proprietary and slowing down innovation in the network. SDN is a new networking approach to solve problems networking and telecommunications is facing today. Though SDN is a framework and not a mechanism and so could mean many different things, it is essentially based on the idea of separating the control plane from the data plane, logically centralizing the control plane and introducing a vendor-agnostic interface between the control and the data plane.

## 2.1 SDN Architecture

As shown in Fig 2(b), the centralized SDN controller maintains a global network map which enables it to take intelligent and globally optimal decisions with respect to routing and switching based on Quality of Service (QoS), traffic load etc. To this effect, the SDN controller implements an abstract forwarding model which comprises of the forwarding function f(Map) which it calculates using the global map information as input.

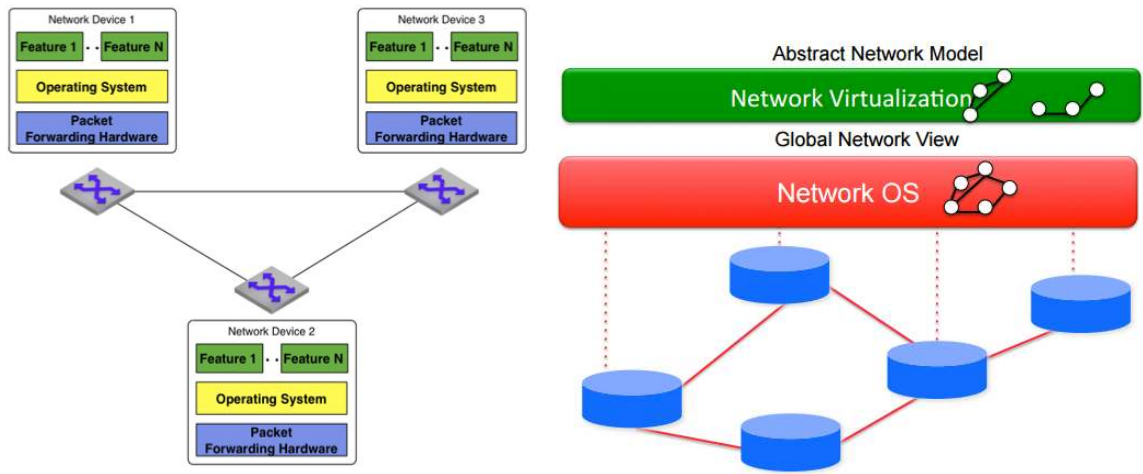

Fig 2: a) Traditional architecture and b) Simplified SDN Architecture [5]

The decision of the controller (say routing in our case) is then communicated to the actual hardware using a secure channel. One of the popular protocols being used for this purpose is OpenFlow.

Fig 3. explains the functioning of OpenFlow, which defines the message format for the communicating the controller decisions to OpenFlow enabled switches and update the flow tables contained within them.

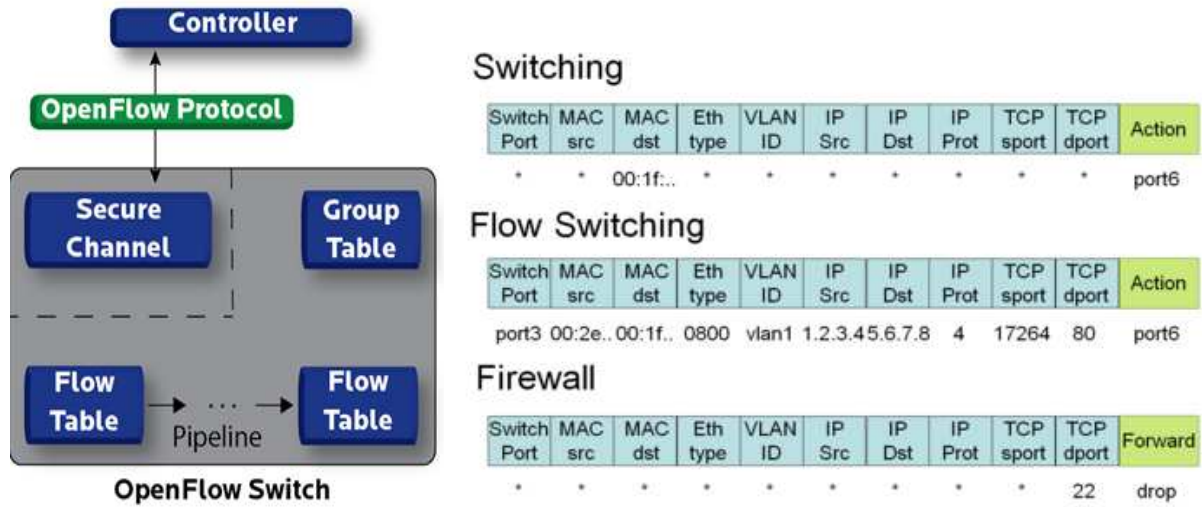

Fig 3: a) Simplified OpenFlow Architecture and b) OpenFlow v1.0 Message format

A flow table basically performs packet lookup. All the arriving packets are compared to flow table entries for a match based on the 12-tuple shown in Fig. The action of the switch depends on the particular match found.

Actions include forwarding the packet, or dropping it, sending it to the controller, modifying it or enqueuing it. Each of these actions have multiple options eg. Forwarding could be LOCAL which means forward to the switch's local networking stack or ALL which means send out on all interfaces except the incoming interfaces.

 If no match is found, the packet is sent to the controller. Fig 4. shows that in a more recent version of OpenFlow (OpenFlow v.1.3), a set of actions may be included in an action set, many of which may in turn be included into a Group to process matching flows.

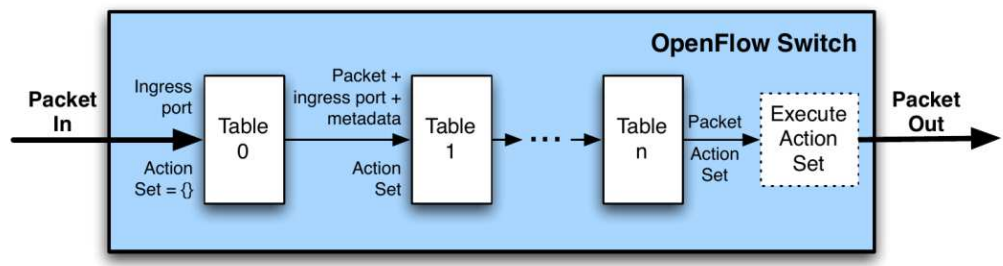

Fig 4: OpenFlow v.1.3 Enhancements: Action Sets and Groups.

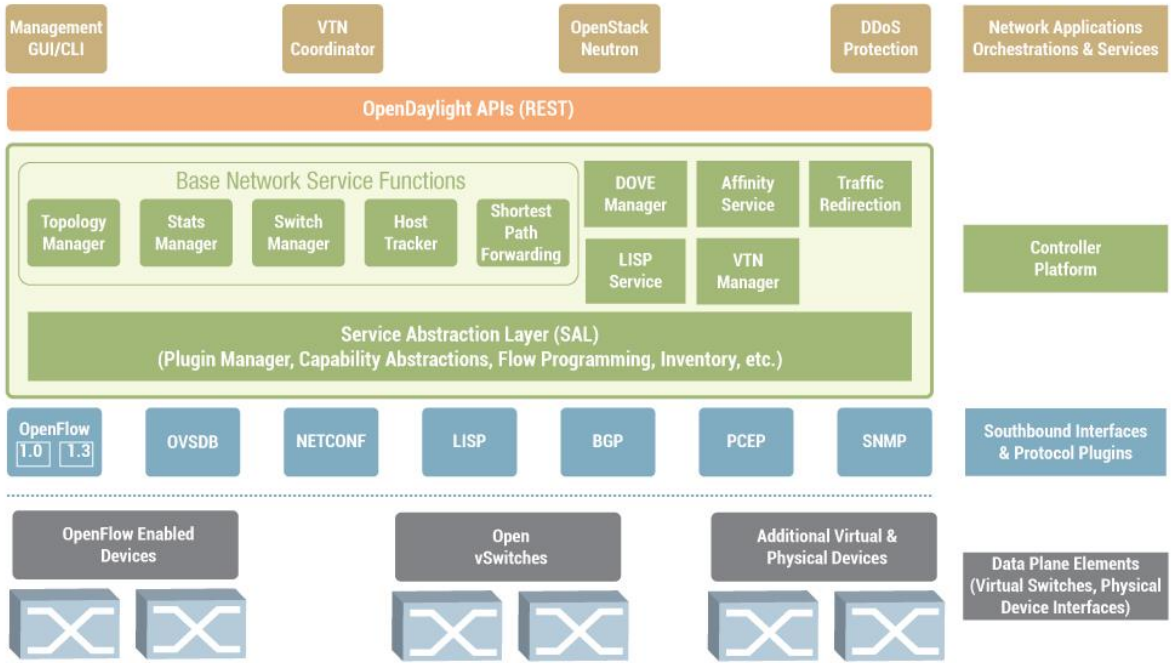

Fig 5: Hydrogen Release of OpenDaylight controller

Fig 5. shows a commercial SDN controller OpenDaylight and how it complies with the SDN framework and also how OpenFlow fits into the architecture.

**2.2 Features of SDN**

We can enumerate the features of SDN as [6]:

  i. *Virtualization:* Use network resource without worrying about where it is physically located, how much it is, how it is organized, etc.
 ii. *Orchestration:* Should be able to control and manage thousands of devices with one command.
iii. *Programmable:* Able to change behavior on the fly.
 iv. *Dynamic Scaling:* Should be able to change size, quantity
  v. *Automation:* To lower OpEx minimize manual involvement i.e. troubleshooting, reduce downtime, Policy enforcement, Provisioning/Re-provisioning/Segmentation of resources, add new workloads, sites, devices, and resources etc.

vi. *Visibility:* Monitor resources, connectivity, performance
vii. *Performance:* Optimize network device utilization, Traffic engineering/Bandwidth management, Capacity optimization, Load balancing, High utilization, Fast failure handling etc.
viii. *Multi-tenancy:* Can have multiple tenants using slicing. Tenants need complete control over their addresses, topology, and routing, security
ix. *Service Integration:* Load balancers, firewalls, Intrusion Detection Systems (IDS), provisioned on demand and placed appropriately on the traffic path
x. *Openness:* Modular plug-ins, Abstractions (API's). Emphasizing on 'what' has to be done and not 'how' it has to be done.

## 3 Evolution of Optical Networks and GMPLS

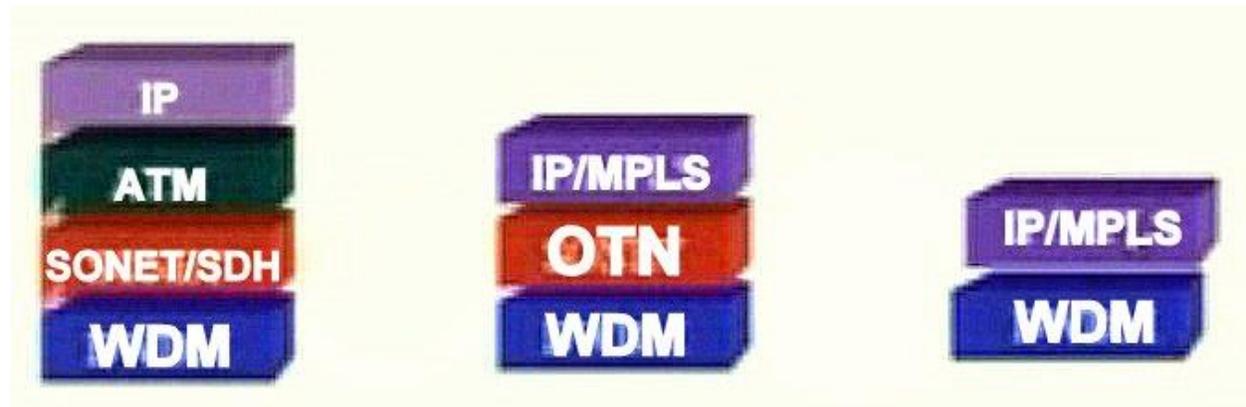

Fig 6: Evolution of Optical Networks

Optical network architecture has undergone a lot of changes over the years. As a result we have legacy architectures coexisting with newer architectures. Initially we had the 4 layer architecture shown in Fig 6. where the ATM layer was basically used for most of the transport functionalities and integration of multiple services. It also provided traffic engineering and QOS capabilities. The SONET or SDH was used to provide multiplexing, protection, restoration capabilities and you could have DWDM at layer 0 to increase the capacity. IP was run over all this to provide internet services.

Historically MPLS was proposed to address the challenges of IP forwarding based on longest prefix match. But it was found that this label based forwarding paradigm could also be used to address the traffic engineering challenges in the IP networks. IP has its own QoS model in terms of integrated services and differentiated services. It was found that MPLS actually helps that IP QoS model. So, by extending this differentiated services and integrated services paradigm to MPLS based network, we could actually provide the quality of service guarantees or could setup the guaranteed bandwidth label switch path or virtual circuit in the core of the networks. So, since the IP/MPLS layer could perform the functions which the ATM layer did, and that too without segmentation and reassembly by acting on variable length IP packets, we did away with the ATM. Also OTN became a protocol of choice in place of SONET/SDH because of its better manageability and also transparency in the sense that it not only allowed mixing of synchronous signals with different timings but also allowed mixing of asynchronous signals. There were also deployments where the SONET/SDH layer was done away with and IP/MPLS now ran in a native mode over this DWDM network. Basically IP ran directly over optical networks.

There are 2 models to consider when talking about IP-Optical converged networks- the overlay model and the Peer model as shown in Fig 7. In the overlay model, the IP network is overlaid over the optical network which comprises of the wavelength routers and separate control planes for the IP network and Optical network. The Overlay model can be further classified into two- Static Overlay model where the lightpaths are statically provisioned and a Dynamic Overlay model in which there is dynamic wavelength provisioning using wavelength routers (WR). Also, in this case the wavelength routers are Optical Cross Connects (OXC's) with IP routing intelligence. In case of Static Overlay networks, the protection and restoration can be done at the IP layer but it is very slow and so we need to leverage the MPLS traffic engineering capabilities to provide protection and restoration. In case of Dynamic overlay model, there is dynamic establishment of light paths which includes features like neighbor discovery, links state update, route computations and the path establishment etc. The best control plane for the wavelength routers turned out to be MPLS. Since the IP network already consisted of IP/MPLS routers, this led to an integrated peer to peer network model for IP-optical converged networks. So, in the Peer model, there is only one control plane for IP routing and optical routing and it spans an entire administrative domain. It must be borne in mind that we had to make some changes in MPLS because it was now dealing with OXC's and not IP packets.

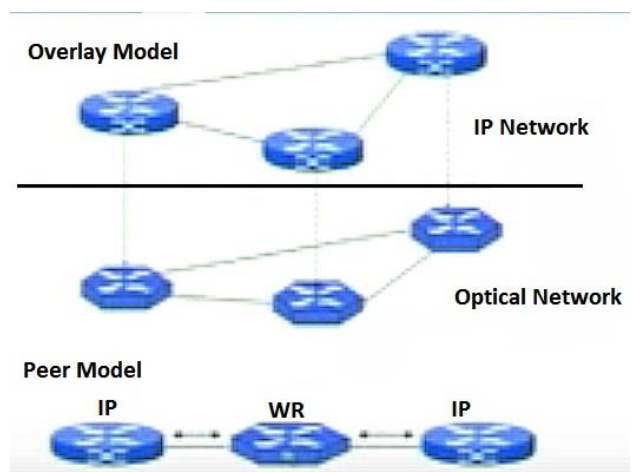

Fig 7: Overlay model and Peer model for IP-Optical Networks

Thus was born the Generalized Multiprotocol Label Switching protocol (GMPLS) or the multiprotocol lambda switching (MPLmS) protocol.

In GMPLS, the edge LSR, provides virtual point to point unidirectional LSP and the wavelength router provides the point to point light path. The LSR maintains the next hop label forwarding entries and the wavelength router (WR) maintains the cross connect table. The WR toggles wavelength from an input port to an output port and a label switch router does label swapping.

So the way this works is, first light paths are determined and distributed using some protocol like OSPF. Then using traffic engineering criteria, path from one node to another node can be determined by using some constrained based routing algorithm. At the edge of the network, all the IP flows which now need to be forwarded onto the optical networks are aggregated into traffic flows , which are aggregated onto traffic trunks which are mapped to the light paths and then the traffic can be switched in the optical cross connect. [7]

The multiprotocol lambda switch router switches the wavelength and not the labels which are there as headers in the optical packets. So, a label here is analogous to a wavelength. In case of a standard LSR, the label is 20 bit and therefore between 2 MPLS nodes we could have one million label switch paths But in optical networks we cannot have such large number of wavelengths and so the granularity of a FEC in the GMPLS node is very course where we map a traffic trunk to a lightpath.

### 3.1 Problems with GMPLS

Some problems in GMPLS become evident immediately. We cannot actually have any hierarchy/ aggregation of these LSP's into a bigger LSP. GMPLS requires extensions to routing protocol being used (OSPF in our case) to be able consider optical factors like bandwidth on the wavelength, optical fiber dispersion and attenuation characteristics etc. GMPLS also required extensions to signaling protocols like CR- LDP or RSVP-TE which are used to convey the traffic trunk attributes to the label switch routers and also requires a link management protocol [7]. Moreover, because of the distributed nature of GMPLS, it has the disadvantages of slow convergence and using limited and outdated information. Another problem attributed to its distributed nature is that global optimization and coordination is very difficult to achieve. Since GMPLS standardization has not happened as rapidly as product development, vendors have devised proprietary extensions mentioned earlier. This has led to production and deployment of GMPLS devices which cannot interwork. Another issue GMPLS faces is that of abstraction. Node resources modelling and control is considered to be an out of scope problem for GMPLS. Additionally, GMPLS software implementations are not based on open source and are the software stack itself is sold bundled with the transport network node. This has resulted in its limited adoption and its practicality still remains debatable [8].

### 4 Software defined Optical Networks (SDONs)

The needs of the network operators, the indispensability of optical transport, benefits of using SDN and the problems with current packet optical solutions point toward bringing the SDN paradigm into optical networks. Hence, on the lines of the aforementioned SDN architecture, a SDON (SDON) architecture could be defined as shown in fig 8. It basically comprises of physical layer, a controller and an open interface to connect the two.

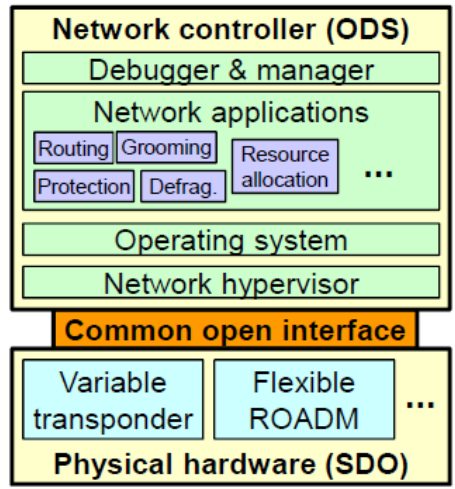

Fig 8: Example Architecture of SDON

**4.1 SDON Physical Hardware or Software defined Optics**

The lowest layer would be the new physical hardware or Software defined Optics which would help create an elastic optical layer which SDN could fully exploit. In [10] it has been proposed to equip the physical layer with a variable transponder or an elastic transponder or a tunable transceiver for transmission. It allows the change of signal characteristics like data rate, modulation format, error correction and coding scheme for the WDM channels taking into account the instantaneous link conditions and QoS requirements. Such a transponder could be achieved in the form of a digital transmitter wherein the data would be encoded using a suitable modulation format using high speed DSP and then modulated onto I and Q phase components of the optical signal through digital-to-analog converters (DAC) and electro-optic modulators. Besides changing the modulation format on the fly, signal can be manipulated beyond standard constellations to achieve effects like reducing crosstalk but still achieving dense packing [11,12]. This approach also allows usage of different DSP-based impairment compensation schemes and selects optimal forward error correction (FEC) coding scheme [13]. But this approach is not power efficient due to high speed electronics used.

Another scheme for variable format transmitter is based on cascaded optical modulators and an electrical-optical- electrical multilevel drive signal generator [14]. Yet another scheme is the variable transmitter scheme is based on all-optical orthogonal frequency division multiplexing (AO-OFDM) [4]. Because in this approach, the transmission performance is not affected by tuning, this approach can be used in conjunction with the first 2 approaches to enhance their flexibility.

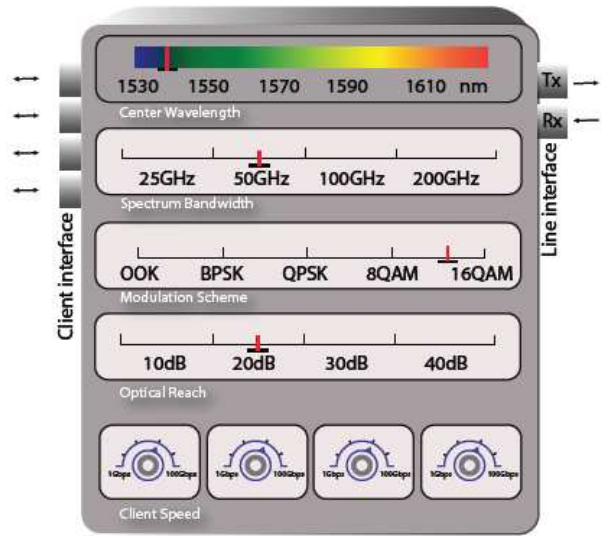

Fig 9: A model of a software-programmable transponder or a tunable transmitter where transmission parameters can be flexibly adjusted based on network demand [10]

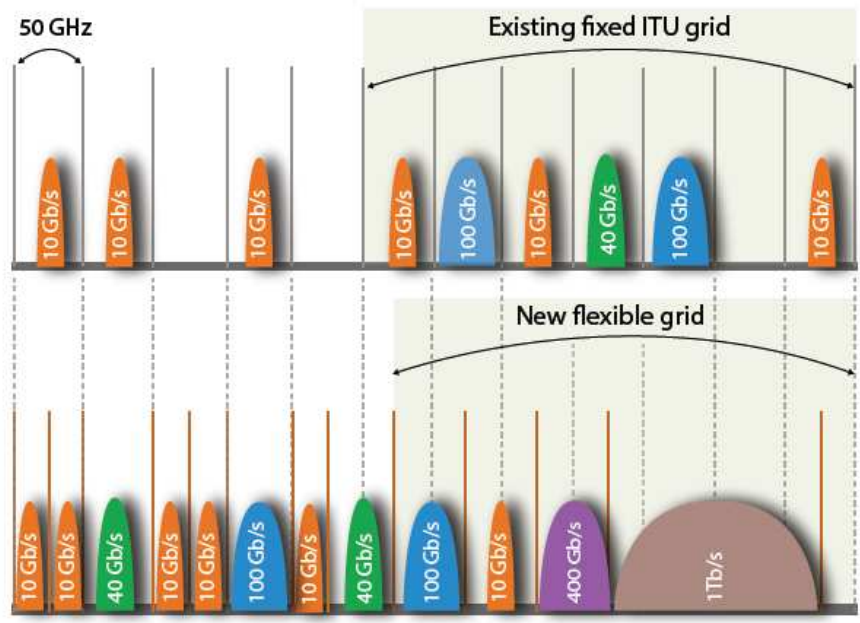

Fig 10: A flexible wavelength grid allows more efficient use of bandwidth for a wider variety of signal bandwidths [10]

The other function of the physical layer is switching which would be accomplished by a flexible switching node. A switching node in today's WDM networks is a ROADM which provides cross connections, add/drop functionality for WDM channels and also performs signal balancing and monitoring. In addition to this, the SDON ROADM would need to provide flexible, non-blocking switching at low cost and without high signal penalty. To be able to this, the flexible ROADM would have to be:

Colorless, Directionless, Contentionless, Gridless meaning non-uniform channel would have to be supported and individual passband width could be changed dynamically to fit signals from variable transponders, Filterless meaning without using a tunable filter or optical demux it would have to be able dynamically select the target flexible grid WDM channel from multiple channels, and Gapless meaning it would have the ability to group contiguous WDM channels headed for same destination into wavebands which could be switched together. This type of a ROADM could be built using DMD or LCoS based wavelength selective switches (WSSes) with programmable passband widths [10, 17, 18, 19]. An additional feature of the WSS's would be that they would be able to function on flexible grids shown in fig 10. which would be needed to accommodate the future bandwidth heterogeneity given that missed-line-rate signals (where different line rates operate over different wavelength channels) are expected to co-exist on the same fiber. A coherent receiver with the local oscillator tuned to the target channel's wavelength can be used to implement the filterless feature.

## 4.2 The SDON Controller

The controller part of the SDON architecture shown in fig 8. would have different specialized networking applications for the optical layer like:

i. *Context aware routing and lightpath selection*.[20,21]
ii. *Network defragmentation:* Due to continuously evolving traffic, the previously optimal control plane may not be optimal all the time and so periodic network defragmentation will help optimize resource utilization.[22]
iii. *Traffic grooming*: Traffic grooming has been demonstrated on the optical layer which improves channel utilization and increases spectral efficiency without involving client layer operations.[23]
iv. *Resource allocation:* The SDON controller can intelligently plan and allocate hardware resources like amplifiers, 2R/3R regenerators and wavelength convertors thus achieving both performance and cost optimization.[24,25]
v. *Protection:* The SDON controller being centralized can easily manage protection and restoration.[26]

## 4.3 Openflow

The open interface mentioned in fig 8. could be OpenFlow whose flexibility, simplicity and manageability make it an attractive solution to the problems of GMPLS integration.

### 4.3.1 Issues in extending OpenFlow to Optical Networks:

Although studies have been conducted, basic extensions for circuit control have not been integrated into the OpenFlow specification. OpenFlow is based on Match/Action Table model of the switch and does not model the internal characteristics, such as port-to port wavelength connectivity limitations in ROADMs and in short, focusses on the actions taken inside the switch and not on connectivity between switches. Path computation needs to consider the impact of optical impairments, end-to-end signal-to-noise ratio etc. The controller also has to inform the receiver of the modulation type, power level, FEC coding, etc. being used at the variable transponder transmitter end, which must be set by application codes. The information is defined by characteristics such as port, timeslot, and wavelength which define the data stream and not the information carried within the data stream such as packet header fields. So the controller cannot request actions based on an analysis of the data. This limits the controller functionality. Optical networks under network operators span large geographical distances, which means the optical network would have different vendor equipment and multiple domains, for which a single controller would be insufficient. Multiple controllers could be employed in such scenarios but the solution of hierarchical controllers has limitations and there are no good East-West (Controller-Controller) interfaces available. Also due to the large geographical distances, control messages would require a time comparable to the 50ms protection and restoration time limit, which would make it necessary to have local, highly available and redundant controllers with fast failover time.

### 4.3.2 How to deploy OpenFlow in optical networks

Although most of the issues discussed above remain to be solved mainly because of the limited support OpenFlow has for optical networks, there have been successful research trials of SDON's.

*OpenFlow Overlay with Transport Tunnels for POI:* OpenFlow supports L2 and above and also the concept of virtual ports. The idea is that when applying OpenFlow to optical networks as an overlay, the endpoints of the configured packet or circuit tunnel. The optical circuit would controlled by separate management system may be a distributed control plane or a path computation element (PCE). The traffic passes through these tunnels i.e. through intermediate switches without visibility to the controller, and traffic engineering through the core is done independently.

Initially, where OpenFlow extensions were focused on the treatment of wavelength, timeslot, etc. as additional match fields as shown in fig 11. a recent research effort in packet optical interconnection uses the concept of logical ports introduced in the recent OpenFlow releases. A physical port on the switch may have multiple logical ports associated with it and each logical port would have characteristics like wavelength, timeslot etc. which can be modified by the controller. This idea is shown in fig 12.

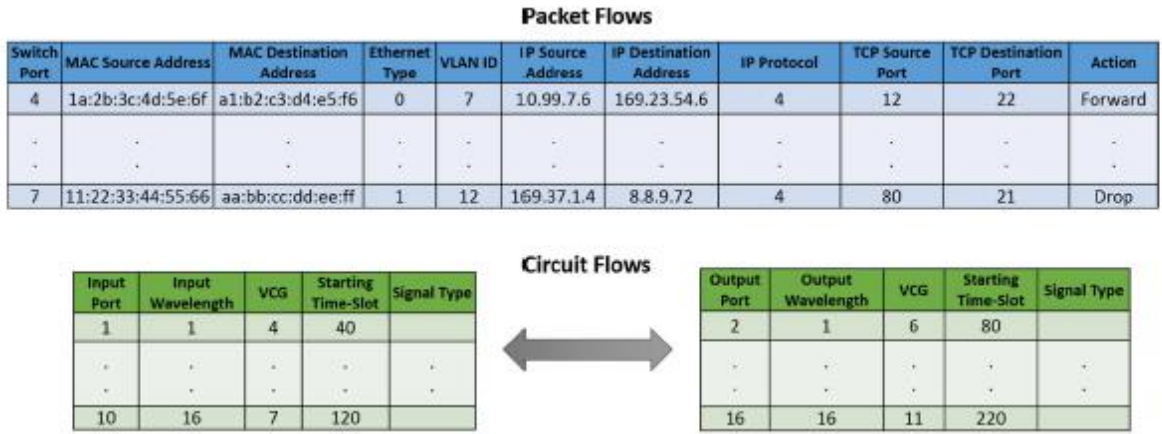

Fig 11. Circuit flows and Packet flows in OpenFlow extension.

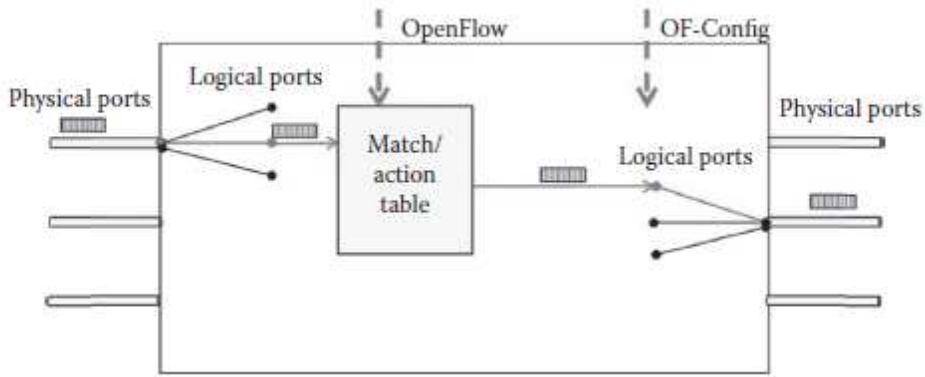

Fig 12: OpenFlow POI model.

*OpenFlow Overlay Over the Control Plane (with Abstraction):* Because it has extensions for the control of circuit mapping and packet forwarding, OpenFlow can be used directly for control over optical switching elements. If there are multiple domains of deployed equipment, OpenFlow could be deployed as an overlay to reduce deployment costs and speed up deployments. The controller would communicate with the EMS via OpenFlow but the control of resources themselves would be managed separately by an intermediate system that would provide an abstract model of a possibly sliced network to the controller as shown in fig 13.

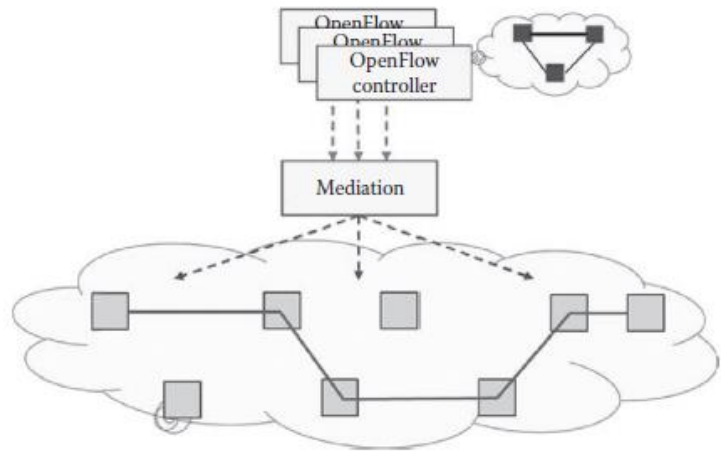

Fig 13: OpenFlow overlay with abstraction.

*Direct OpenFlow Control of Packet/Optical:* This approach involves introducing an OpenFlow agent on every network element and adding connectivity from each network element to the controller as shown in fig 14.

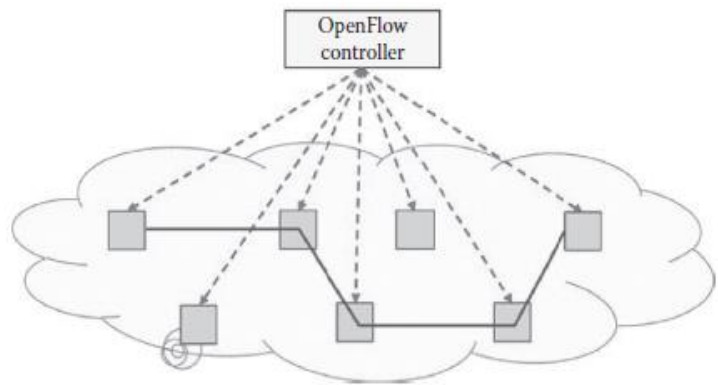

Fig 14: OpenFlow direct/centralized control

### 4.4 Coexistence of GMPLS and OpenFlow

Until OpenFlow is ripe enough to handle the entire unified control plane by itself, coexistence of GMPLS and OpenFlow could be used as a stopgap measures could be employed. Let us briefly see 2 of the many models that have been proposed to this effect.

*OpenFlow over GMPLS:* In fig. 15 we can see that GMPLS groups a set of nodes (eg. a WDM ring) and exports an OF virtual node. The GMPLS northbound interface should be OF node like.

*GMPLS over OpenFlow:* In fig. 16 we can see that in this case, OpenFlow becomes the standard for GMPLS southbound interface for all nodes and it would require implementing a 'virtual' GMPLS topology across OpenFlow controlled switching domains.

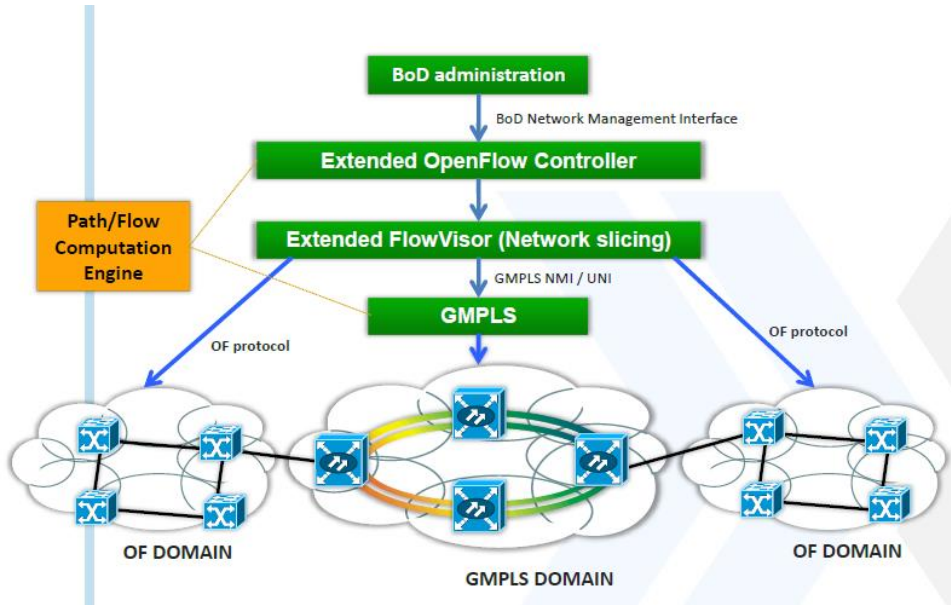

Fig 15: OpenFlow over GMPLS

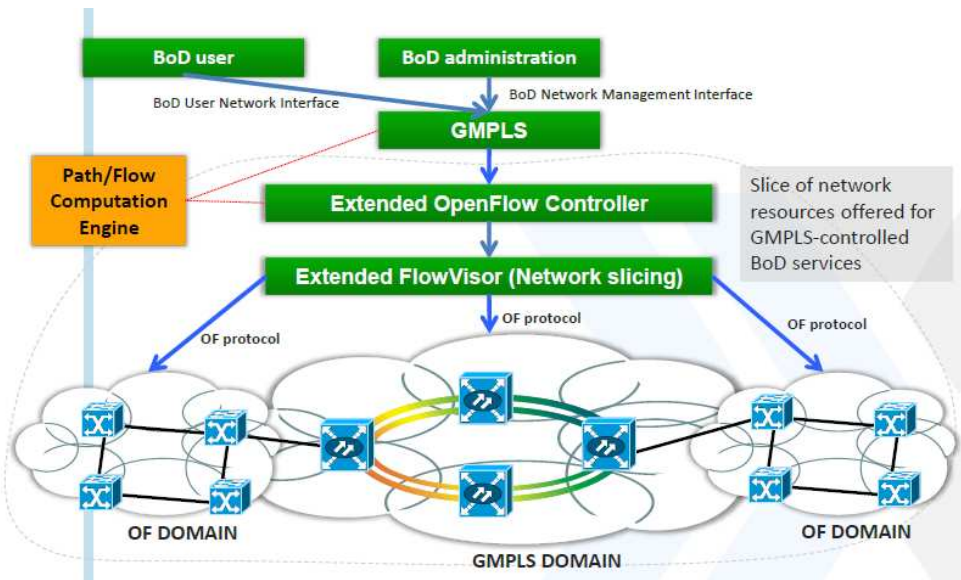

Fig 16: GMPLS over OpenFlow

## 5 Conclusion

It is clear how SDON can bring multiple benefits to the way we do optical communications. It is also clear that combining the SDN paradigm with optical networks is the way to go to build intelligent, high capacity transport networks which the applications of today demand. Building a deployable SDON solution requires extensive research and innovation not only in the hardware, controller and Openflow but also in the type of network applications we build, the business models which govern them and also schemes to achieve cross layer optimization.